\documentclass[12pt]{article}
\usepackage{amssymb,amsmath,amsthm,amsfonts,amscd}
 \usepackage{graphicx}
\newcommand\be{\begin{equation}}
\newcommand\ee{\end{equation}}
\newcommand\bea{\begin{eqnarray}}
\newcommand\eea{\end{eqnarray}}
\newcommand\ket[1]{|#1\rangle}
\newcommand\bra[1]{\langle #1|}

\newcommand{\fatalpha}{{\bf \alpha \kern -0.44em \alpha}}
\newcommand{\fatsigma}{{\bf \sigma \kern -0.54em \sigma}}
\newcommand{\tpchi}{{\bf D \kern -0.35em D}}
\newcommand{\llambda}{{\bf \lambda \kern -0.45em \lambda}}

\renewcommand{\theequation}{\arabic{equation}}
\renewcommand{\theequation}{\thesection.\arabic{equation}}
\bibliography{plain}
\title{\bf Spin-momentum correlation in
relativistic single particle quantum states    } \vspace{20mm}
\author{ M. A. Jafarizadeh$^{a,b,c}$
 \thanks{E-mail:jafarizadeh@tabrizu.ac.ir},
  M. Mahdian$^{a}$
 \thanks{E-mail:Mahdian@tabrizu.ac.ir}\\
$^a${\small Department of Theoretical Physics and Astrophysics,
University of Tabriz, Tabriz 51664, Iran.}  \\ $^b${\small Institute
for Studies in Theoretical Physics and Mathematics, Tehran
19395-1795, Iran.}\\$^c${\small Research Institute for Fundamental
Sciences, Tabriz 51664, Iran.}} \pagebreak

\vspace{20mm}

\begin{document}
\maketitle \vspace{15mm}
\newpage
\begin{abstract}
This paper was concerned with the spin-momentum correlation in
single-particle quantum states, which is described by the mixed
states under Lorentz transformations. For convenience, instead of
using the superposition of momenta we use only two momentum
eigen states ($p_1$ and $p_2$) that are perpendicular to the
Lorentz boost direction. Consequently, in 2D momentum subspace we show
that the entanglement of spin-momentum in the moving frame
depends on the angle between them. Therefore, when spin and momentum are perpendicular the measure of entanglement is not observer-dependent quantity in inertial frame. Likewise, we have calculated the
measure of entanglement (by using the concurrence) and has shown that  entanglement decreases with
respect to the increasing of observer velocity.
Finally, we argue that, Wigner rotation is induced by Lorentz
transformations  can be realized as controlling operator.

\end{abstract}

\section{Introduction}
In two recent decades quantum entanglement has become as  one of the most important resources in the
rapidly growing field of quantum information processing with
remarkable applications on it \cite{Einstein}, and was based on the fact that the
existence of entangled states produces nonclassical
phenomena. Therefore, specifying that a particular quantum state is
entangled or separable is important because if the quantum state be
separable then its statistical properties can be explained entirely
by classical statistics.\par Relativistic aspects of quantum
mechanics have recently attracted much attention in the
context of the theory of quantum information, especially on quantum entanglement\cite{Harshman,Lee,Caban,Bartlett,peres,Wigner,Adami,Terashimo,Czachor,Alsing,Ahn,Jafarizadeh,lamata1,lamata2,Jian,Jason,Andr}.
Peres \emph{et} al.\cite{peres} have recently observed that the reduced spin
density matrix of a single spin-$\frac{1}{2}$ particle is not a
relativistic invariant, and Wigner rotations
correlate spin with the particle momentum distribution when is
observed in a moving frame \cite{Wigner}. Gingrich and Adami
have shown that the entanglement between the spins of two
particles is carried over to the entanglement between the momenta of
the particles by the Wigner rotation, even though the entanglement
of the entire system is Lorentz invariant \cite{Adami}. Terashimo and Ueda
\cite{Terashimo} and Czarchor \cite{Czachor} suggested that the
degree of violation of the Bell inequality depends on the velocity
of the pair of spin-$\frac{1}{2}$ particles or the observer with
respect to the laboratory. Alsing and Milburn  studied
the Lorentz invariance of entanglement and showed that the
entanglement fidelity of the bipartite state is preserved
explicitly. Instead of state vector in the Hilbert space, they
have used a 4-component Dirac spinor or a polarization vector in
favor of quantum field theory \cite{Alsing}. Ahn  also calculated the
degree of violation of the Bell's inequality which is decreases with
increasing of velocity of the observer \cite{Ahn}. Most of the previous works
were concerned with the  pure states although authors in
\cite{Jafarizadeh,lamata1,lamata2} have considered mixed quantum states that are described by  superposition of momenta with
Gaussian  distribution, where  Lorentz transformation introduces a
transfer of entanglement between different degrees of freedom. While
the entanglement between spins and momentums of particles  may change, separately. However, the total  entanglement of particle-particle is the same in
all inertial frames. Beside of the pervious works that are concerned of study on the entanglement between quantum states of two particles, here we generalize this to the spin-momentum correlation of relativistic single-particle (by using the concurrence) and show that the measure of entanglement depends on the  angle between spin and momentum and it decreases with increasing of velocity of the observer. Also it has been shown that the Wigner
angle depends on momentum, so Wigner rotation behaves as a quantum gate or controlling operators. Thus using this quantum gate the
spin-momentum entanglement changes in the framework of
special relativity.\\
This paper is organized as follows:  Sec. II, is devoted
to single-particle relativistic quantum states. In Sec. III, we calculate
explicitly the spin-momentum entanglement of relativistic quantum state. In sec. IV, we
explain how we can use the quantum gate via Lorentz transformation.
The last section contains the concluding remarks. The paper also
contains two appendixes.

\section{Single-particle relativistic quantum states}

Suppose we have a bipartite system with its quantum degrees of
freedom distributed among two parties $\mathcal{A}$ and
$\mathcal{B}$ with Hilbert spaces $\mathcal{H}_A$ and $\mathcal{H}_B$, respectively, (the standard Hilbert space of dimension $d$ endowed
with usual inner product denoted by $\langle\ .\ \rangle$). In
this paper quantum state is made up of a single-particle having two
types of degrees of freedom : momentum \emph{p} and spin $\sigma$.
The former is a continuous variable with Hilbert space of infinite
dimension but we restrict ourselves here  to 2D momentum subspace with two eigen-state $p_1$ and $p_2$, while the latter is a discrete one with Hilbert space of spin particle. The pure quantum state of such a system can always be written as
\be\label{p1}\ket{\psi}=\sum_{i=1}^{2}\sum_{j=n}^{-n}
c_{ij}\ket{p_i}\otimes \ket{j},\ee\\
where  $\ket{p_{1(2)}}$  are two momentum eigen states of each particle and the kets $\ket{j}$  are the eigenstates of spin operator. $c_{ij}$'s are complex coefficients such that $\sum
_{i,j}|c_{ij}|^2=1 $ .

A bipartite quantum mixed state is defined as a
convex combination of
 bipartite pure states (\ref{p1}), i.e. \be\label{ro1}\rho=\sum_{i=1}^4 P_i
\ket{\psi_{i}} \bra{\psi_{i}},\ee where $P_i\geq 0$, $\sum _{i} P_i=1 .$ $\ket{\psi_i}$ $(i=1,2,3,4)$ as four orthogonal maximal entangled Bell states (\emph{BD}) are belong to
 the product space $\mathcal{H}_A \otimes \mathcal{H}_B$ and  in terms of momentum and spin states are well-known as
$$ \ket{\psi_1}=\frac{1}{\sqrt{2}}(\ket{p_{1}}\otimes\ket{n}+\ket{p_{2}}\otimes\ket{-n}) ,$$
$$ \ket{\psi_2}=\frac{1}{\sqrt{2}}(\ket{p_{1}}\otimes\ket{n}-\ket{p_{2}}\otimes\ket{-n}) ,$$
$$ \ket{\psi_3}=\frac{1}{\sqrt{2}}(\ket{p_{2}}\otimes\ket{n}+\ket{p_{1}}\otimes\ket{-n}) ,$$
\be\label{pure2}\ket{\psi_4}=\frac{1}{\sqrt{2}}(\ket{p_{2}}\otimes\ket{n}-\ket{p_{1}}\otimes\ket{-n}).\ee\\
Here, $\ket{\pm n}$ are the Bloch sphere representation of  spin state (qubit) as
\be\label{spin}\ket{n}=\left(\begin{array}{c}\cos{\frac{\xi}{2}} \\
e ^{i\tau} \sin{\frac{\xi}{2}}\end{array}\right) ,  \ket{-n}=\left(\begin{array}{c}\ \sin{\frac{\xi}{2}} \\
-e ^{i\tau}  \cos{\frac{\xi}{2}}\end{array}\right), \ee
where $\xi$ and $\tau$ are polar and azimuthal  angles, respectively.
\subsection{Relativistic  single spin-$\frac{1}{2}$ particle quantum states}
We assumed that spin and momentum are in the yz-plane ( $\tau=\frac{\pi}{2}$ , $\vec{p}=(0,
p\sin{\theta}, p\cos{\theta})) $ and the Lorentz boost is orthogonal to it. For an observer in another
reference frame $S^\prime$ described by an arbitrary boost $\Lambda$
in the x-direction, the transformed \emph{BD} states are given by (see Appendix A)
$$\ket{\psi_{i}}\longrightarrow U(\Lambda)\ket{\psi_{i}} ,$$\
$$\vspace{-6mm}\hspace{-60mm}\ket{\Lambda\psi_1}=\frac{1}{\sqrt{2}}\left\{\ket{\Lambda
p_{1}}\otimes \left(\begin{array}{c}\ \cos{\frac{\xi}{2}}\cos{\frac{\Omega_{\vec{p_{1}}}}{2}}-i\sin{\frac{\Omega_{\vec{p_{1}}}}{2}}\sin{\frac{\zeta}{2}} \\
i\sin{\frac{\xi}{2}}\cos{\frac{\Omega_{\vec{p_{1}}}}{2}}+\sin{\frac{\Omega_{\vec{p_{1}}}}{2}}\cos{\frac{\zeta}{2}}\end{array}\right)\right.$$
$$\hspace{95mm}\left.+\ket{\Lambda
p_{2}}\otimes \left(\begin{array}{c}\ \sin{\frac{\xi}{2}}\cos{\frac{\Omega_{\vec{p_{2}}}}{2}}-i\sin{\frac{\Omega_{\vec{p_{1}}}}{2}}\cos{\frac{\zeta}{2}} \\
-i\cos{\frac{\xi}{2}}\cos{\frac{\Omega_{\vec{p_{2}}}}{2}}-\sin{\frac{\Omega_{\vec{p_{2}}}}{2}}\sin{\frac{\zeta}{2}}\end{array}\right)\right\},$$
$$\vspace{-6mm}\hspace{-60mm} \ket{\Lambda\psi_2}=\frac{1}{\sqrt{2}}\{\ket{\Lambda
p_{1}}\otimes \left(\begin{array}{c}\ \cos{\frac{\xi}{2}}\cos{\frac{\Omega_{\vec{p_{1}}}}{2}}-i\sin{\frac{\Omega_{\vec{p_{1}}}}{2}}\sin{\frac{\zeta}{2}} \\
i\sin{\frac{\xi}{2}}\cos{\frac{\Omega_{\vec{p_{1}}}}{2}}+\sin{\frac{\Omega_{\vec{p_{1}}}}{2}}\cos{\frac{\zeta}{2}}\end{array}\right)$$
$$\hspace{95mm}\left.-\ket{\Lambda
p_{2}}\otimes \left(\begin{array}{c}\ \sin{\frac{\xi}{2}}\cos{\frac{\Omega_{\vec{p_{2}}}}{2}}-i\sin{\frac{\Omega_{\vec{p_{1}}}}{2}}\cos{\frac{\zeta}{2}} \\
-i\cos{\frac{\xi}{2}}\cos{\frac{\Omega_{\vec{p_{2}}}}{2}}-\sin{\frac{\Omega_{\vec{p_{2}}}}{2}}\sin{\frac{\zeta}{2}}\end{array}\right)\right\},$$
$$ \vspace{-6mm}\hspace{-60mm}\ket{\Lambda\psi_3}=\frac{1}{\sqrt{2}}\{\ket{\Lambda
p_{2}}\otimes \left(\begin{array}{c}\ \cos{\frac{\xi}{2}}\cos{\frac{\Omega_{\vec{p_{2}}}}{2}}+i\sin{\frac{\Omega_{\vec{p_{2}}}}{2}}\sin{\frac{\zeta}{2}} \\
i\sin{\frac{\xi}{2}}\cos{\frac{\Omega_{\vec{p_{2}}}}{2}}-\sin{\frac{\Omega_{\vec{p_{2}}}}{2}}\cos{\frac{\zeta}{2}}\end{array}\right)$$

$$\hspace{95mm}\left.+\ket{\Lambda
p_{1}}\otimes \left(\begin{array}{c}\ \sin{\frac{\xi}{2}}\cos{\frac{\Omega_{\vec{p_{1}}}}{2}}+i\sin{\frac{\Omega_{\vec{p_{1}}}}{2}}\cos{\frac{\zeta}{2}} \\
-i\cos{\frac{\xi}{2}}\cos{\frac{\Omega_{\vec{p_{1}}}}{2}}+\sin{\frac{\Omega_{\vec{p_{1}}}}{2}}\sin{\frac{\zeta}{2}}\end{array}\right)\right\},$$

\be\label{rpure}
\vspace{-6mm}\hspace{-60mm}\ket{\Lambda\psi_4}=\frac{1}{\sqrt{2}}\{\ket{\Lambda
p_{2}}\otimes \left(\begin{array}{c}\ \cos{\frac{\xi}{2}}\cos{\frac{\Omega_{\vec{p_{2}}}}{2}}+i\sin{\frac{\Omega_{\vec{p_{2}}}}{2}}\sin{\frac{\zeta}{2}} \\
i\sin{\frac{\xi}{2}}\cos{\frac{\Omega_{\vec{p_{2}}}}{2}}-\sin{\frac{\Omega_{\vec{p_{2}}}}{2}}\cos{\frac{\zeta}{2}}\end{array}\right)$$

$$\hspace{95mm}\left.-\ket{\Lambda
p_{1}}\otimes \left(\begin{array}{c}\ \sin{\frac{\xi}{2}}\cos{\frac{\Omega_{\vec{p_{1}}}}{2}}+i\sin{\frac{\Omega_{\vec{p_{1}}}}{2}}\cos{\frac{\zeta}{2}} \\
-i\cos{\frac{\xi}{2}}\cos{\frac{\Omega_{\vec{p_{1}}}}{2}}+\sin{\frac{\Omega_{\vec{p_{1}}}}{2}}\sin{\frac{\zeta}{2}}\end{array}\right)\right\},\ee\\
where $\zeta=(\xi-2\theta)$ and $\{\ket{\Lambda p_{1}},\ket{\Lambda p_{2}}\}$ are two orthogonal
 momentum eigen-state after Lorentz transformation.\\
 The \emph{BD} density matrix (\ref{ro1}), which describes the state of the single-particle at non-relativistic frame, is exchanged to the density matrix $\rho^\prime$ after Lorentz transformation, i.e.
$$\rho\longrightarrow U(\Lambda)\rho ,$$
\be \label{rro}\rho^\prime=U(\Lambda)\rho=\sum_{i=1}^{4}
P_{i}\ket{\Lambda \psi_{i}}\bra{\Lambda \psi_{i}} .\ee\\
It can be calculate that $\ket{\psi_{i}}$ will be orthogonal after Lorentz transformation, i. e.
$$\langle\Lambda \psi_{i}\ket{\Lambda \psi_{j}}=\delta_{ij}.$$

\section{Spin-momentum correlation }

We know that a system is  entangle when its density
matrix cannot be written as a convex sum of product states. For a
pure state, dividing the system into two subsystems, $\mathcal{A}$
and $\mathcal{B}$, allows the Von Neumann entropy to be used as a
measure of entanglement that corresponds to Ref.\cite{peres} is not
Lorentz invariant. When a bipartite system is in a mixed state,
there are a number of proposals for measures of the entanglement of
it, including the entanglement of formation
\cite{wootters,Wootters2,chen,lozinski}, relative entropy of
entanglement \cite{Vedral} and distillation of
entanglement\cite{Bennett}. For pure states Each of these reduce to the von Neumann entropy. The most well-know bipartite measure of entanglement is entanglement
of formation. Because of this, we apply the concurrence which is introduced by Wootters  related to the entanglement of
formation to measure the mixed-state entanglement of spin-momentum in the
inertial frame.

\subsection{Spin-momentum correlation of pure state}

We show that by the von Neumann entropy the entanglement for a pure state in the
Schmidt form \cite{nielsen}  is not invariant after Lorentz transformation, and depend on the angles between spin and momentum. We introduce the following
pure state
 \be\label{pure1}\ket{\psi}=\sqrt{\lambda_1}
\ket{n}\otimes\ket{p_1}+\sqrt{\lambda_2} \ket{-n}\otimes\ket{p_2},
\ee where $\lambda_1+\lambda_2=1$.\\
We take the trace over the momentum eigen states and we obtain the following reduced spin density matrix\\
$$\rho^\prime=Tr_{\Lambda p_1,\Lambda p_2} (\ket{\Lambda\psi} \bra{\Lambda\psi}),
$$
with the following two different eigenvalues

$$\eta_1=\frac{1}{2}\{\lambda_1+\lambda_2-\sqrt{\lambda_1^2+\lambda_2^2+\lambda_1\lambda_2(\cos{2\varphi}-2\cos^2{\varphi}\cos{(\Omega_{p_1}-\Omega_{p_2})}-1)}\},$$
\be\label{eig1}
\eta_2=\frac{1}{2}\{\lambda_1+\lambda_2+\sqrt{\lambda_1^2+\lambda_2^2+\lambda_1\lambda_2(\cos{2\varphi}-2\cos^2{\varphi}\cos{(\Omega_{p_1}-\Omega_{p_2})}-1)}\},\ee
where $\varphi$ is the angle between the
spin and momentum ( $\varphi=\xi-\theta$).
After some mathematical manipulations we have( see Appendix B)
\be\label{ant1}E(\rho^\prime)\leq E(\rho). \ee
It shows that inequality (\ref{ant1}) shows that when Lorentz boost and momentum are perpendicular, spin-momentum entanglement is decreases with increasing
of velocity of the observer, as well as when spin and momentum are
perpendicular, i. e.
$$\varphi=\frac{\pi}{2}\Rightarrow \eta_1=\lambda_1 , \eta_2=\lambda_2, $$
that show  Lorentz transformation does not change the entanglement
between them, i.e.
$E(\rho^\prime)= E(\rho).$

\subsection{Spin-momentum entanglement of  mixed state}

This subsection is devoted to calculate the concurrence of relativistic \emph{BD} mixed
state is given in (\ref{rro}). By using the Appendix A, we obtain the following result:

$$\lambda_1=\frac{1}{2\sqrt{2}}\{\sqrt{A_1+B_1-\sqrt{C_1 D_1}}\},$$
$$\lambda_2=\frac{1}{2\sqrt{2}}\{\sqrt{A_1+B_1+\sqrt{C_1 D_1}}\},$$
$$\lambda_3=\frac{1}{2\sqrt{2}}\{\sqrt{A_2+B_2-\sqrt{C_2 D_2}}\},$$
$$\lambda_4=\frac{1}{2\sqrt{2}}\{\sqrt{A_2+B_2+\sqrt{C_2 D_2}}\},$$
where
$$\hspace{-22mm}A_{1(2)}=3P_{2(1)}^2+3P_{3(4)}^2-(P_{2(1)}^2+P_{3(4)}^2)\cos{2\varphi},$$
$$\hspace{-14mm}B_{1(2)}=2\cos^2{\varphi}(2P_{2(1)} P_{3(4)}+(P_{2(1)}-P_{3(4)})^2\cos{\omega}),$$
$$\hspace{-12mm}C_{1(2)}=(P_{2(1)}-P_{3(4)})^2(-3+\cos{2\varphi}-2\cos^2{\varphi}\cos{\omega}),$$
$$\hspace{31mm}D_{1(2)}=(-(3P_{2(1)}+P_{3(4)})(P_{2(1)}+3P_{3(4)})+(P_{2(1)}-P_{3(4)})^2(\cos{2\varphi}-2\cos^2{\varphi}\cos{\omega}))
,$$\\
where $\lambda_i$'s are the square roots of the eigenvalues $\rho
\tilde{\rho}$ and $\omega=(\Omega_{p_1}+\Omega_{p_2}).$  First index "1" in
(A,B,C,D) corresponds to the $(P_2,P_3)$ and the second index "2"
corresponds to the $(P_1,P_4),$\\
Therefore
\be\label{con1}C(\rho^\prime)=max\{0,\lambda_1-\lambda_2-\lambda_3-\lambda_4\}
,(\lambda_1\geq \lambda_2\geq \lambda_3 \geq \lambda_4 ).\ee
To see the behavior of concurrence with respect to the boost in x
direction, after some
calculation we obtain the following results,
$$\hspace{-14mm}(\lambda_{1(3)}-\lambda_{2(4)})^2=(P_{2(1)}-P_{3(4)})^2(1-\cos^2{\varphi}\sin^2{\frac{\omega}{2}}),$$
\be\label{eib2}\hspace{8mm}(\lambda_{1(3)}+\lambda_{2(4)})^2=(P_{2(1)}+P_{3(4)})^2-(P_{2(1)}-P_{3(4)})^2\cos^2{\varphi}\sin^2{\frac{\omega}{2}}.\ee\\
Using by Eqs (\ref{eib2}), we obtain\\
$$(\lambda_3-\lambda_4)-(\lambda_1+\lambda_2)=(P_1-P_4)\sqrt{(1-\cos^2{\varphi}\sin^2{\frac{\omega}{2}})}-\sqrt{(P_2+P_3)^2-(P_2-P_3)^2\cos^2{\varphi}\sin^2{\frac{\omega}{2}}},$$\\
it is easy to see that
$$(P_2+P_3)^2-(P_2-P_3)^2\cos^2{\varphi}\sin^2{\frac{\omega}{2}} \geq
(P_2+P_3)^2 (1-\cos^2{\varphi}\sin^2{\frac{\omega}{2}}),$$
so we have
$$(\lambda_3-\lambda_4)-(\lambda_1+\lambda_2)\leq
(P_1-P_4)\sqrt{(1-\cos^2{\varphi}\sin^2{\frac{\omega}{2}})}
-(P_2+P_3)\sqrt{(1-\cos^2{\varphi}\sin^2{\frac{\omega}{2}})}$$
$$ =(P_1-P_4-P_2-P_3)\sqrt{(1-\cos^2{\varphi}\sin^2{\frac{\omega}{2}})}\leq
(P_1-P_4-P_2-P_3)$$
therefore
$$C(\rho^\prime)\leq C(\rho).
$$\\
This shows that the spin-momentum correlation in single-particle
mixed quantum state is dependent on the angle between spin and momentum. Likewise, when spin and momentum are perpendicular, i. e. $\varphi=\frac{\pi}{2}$ then the concurrence is not an observer-dependent quantity in inertial frame, namely $ C(\rho^\prime)= C(\rho)$.

\section{Manipulating Quantum control gates via Lorentz transformation }

We explain how the Lorentz transformations can be realized  as
quantum control gates. To do this, we consider the pure sate
of (\ref{pure1}) under Lorentz transformations as

\be\label{wi1}U(\Lambda)\ket{\psi}=\sqrt{\lambda_1} \ket{\Lambda
p_1}\otimes W(n_1,p_1)\ket{n_1}+\sqrt{\lambda_2} \ket{\Lambda
p_2}\otimes W(n_2,p_2)\ket{n_2},\ee where $ W(n_i,p_j)$ is Wigner
rotation and the spinors  are rotated by the Wigner angles. As a result, the Wigner rotation
essentially behaves like  a quantum control gate or controlling
operator with the control quantum sates $\{\ket{p_1},\ket{p_2}\}
$ and target states $(\ket{n_1},\ket{n_2})$.
In order to better see the quantum control gate, we assume that the reference frame $S'$
is described by an arbitrary Lorentz boost in the x-direction and momentum
and spin are parallel in  z-direction, i.e. $\varphi=0$. Then the
transformed  states in $2\otimes 2$ Hilbert space are given
by
$$ \ket{p_1}\otimes\ket{\frac{1}{2}}\rightarrow \cos{\frac{\Omega_{p_1}}{2}}
\ket{\Lambda
p_1}\otimes\ket{\frac{1}{2}}+\sin{\frac{\Omega_{p_1}}{2}}\ket{\Lambda
p_1}\otimes\ket{-\frac{1}{2}}, $$
$$\hspace{9mm} \ket{p_1}\otimes\ket{-\frac{1}{2}}\rightarrow -\sin{\frac{\Omega_{p_1}}{2}}
\ket{\Lambda
p_1}\otimes\ket{\frac{1}{2}}+\cos{\frac{\Omega_{p_1}}{2}}\ket{\Lambda
p_1}\otimes\ket{-\frac{1}{2}}, $$
$$\ket{p_2}\otimes\ket{\frac{1}{2}}\rightarrow \cos{\frac{\Omega_{p_2}}{2}}
\ket{\Lambda
p_2}\otimes\ket{\frac{1}{2}}+\sin{\frac{\Omega_{p_2}}{2}}\ket{\Lambda
p_2}\otimes\ket{-\frac{1}{2}}, $$
$$\hspace{9mm} \ket{p_2}\otimes\ket{-\frac{1}{2}}\rightarrow -\sin{\frac{\Omega_{p_2}}{2}}
\ket{\Lambda
p_2}\otimes\ket{\frac{1}{2}}+\cos{\frac{\Omega_{p_2}}{2}}\ket{\Lambda
p_2}\otimes\ket{-\frac{1}{2}}, $$
where $\Lambda$ as  matrix representation of the Lorentz transformation in the
computational basis $\{ \ket{\Lambda
p_1}\ket{\frac{1}{2}},\ket{\Lambda
p_1}\ket{-\frac{1}{2}},\ket{\Lambda
p_2}\ket{\frac{1}{2}},\ket{\Lambda p_2}\ket{-\frac{1}{2}}\}$ is calculated as:
$$
\begin{array}{cccc}
 \Lambda= \left(\begin{array}{cccc}
  \cos{\frac{\Omega_{p_1}}{2}} & \sin{\frac{\Omega_{p_1}}{2}}&0&0\\
-\sin{\frac{\Omega_{p_1}}{2}} & \cos{\frac{\Omega_{p_1}}{2}}&0&0\\
0 & 0&\cos{\frac{\Omega_{p_2}}{2}}&\sin{\frac{\Omega_{p_2}}{2}}\\
0 & 0&-\sin{\frac{\Omega_{p_2}}{2}}&\cos{\frac{\Omega_{p_2}}{2}}\\
\end{array}
\right), \\
\end{array}
$$
In the special case where $(\Omega_{p_2}+\Omega_{p_1})=\pi$, we obtain
$$\cos{\frac{\Omega_{p_2}}{2}}=\sin{\frac{\Omega_{p_1}}{2}},\  \
\sin{\frac{\Omega_{p_2}}{2}}=\cos{\frac{\Omega_{p_1}}{2}}$$ and
 in the limit of $\Omega_{p_1}\rightarrow 0$ we get
 \be\label{mat11}
\begin{array}{cccc}
Lim_{_{_{\hspace{-3mm}\Omega_{p_1}\rightarrow 0}}}\hspace{-6mm}\Lambda=\left(\begin{array}{cccc}
  1 & 0&0&0\\
0 & 1&0&0\\
0 & 0&0&1\\
0 & 0&-1&0\\
\end{array}
\right). \\
\end{array}
\ee
We know that, the Controled-Not (CNOT) gate is a two-qubit circuit that transforms target qubit from its
initial eigen-state to the opposite basis state iff the 'control'
qubit is in eigen-state $\ket{-\frac{1}{2}}$. Obviously, the quantum operation
(\ref{mat11}) flips the spin states, when
the control momentum state is $\ket{p_2}$, so the matrix
representation (\ref{mat11}) is similar to
the Controlled-Not (CNOT) gate. This CNOT is a nonlocal
operation because it can actually create a maximally entangled
state from a product state or vice versa. For instance, after applying
the gate($\ref{mat11}$)on the product
state$(\ket{p_1}+\ket{p_2})\otimes\ket{\frac{1}{2}}$, we obtain
the following entangled state
\be\label{en1}(\ket{p_1}+\ket{p_2})\otimes\ket{\frac{1}{2}}\rightarrow
\ket{\Lambda p_1}\otimes\ket{\frac{1}{2}} +\ket{\Lambda
p_2}\otimes\ket{-\frac{1}{2}},\ee
and for maximally-entangled Bell state
\be\label{en2}
\frac{1}{\sqrt{2}}(\ket{p_{1}}\otimes\ket{\frac{1}{2}}+\ket{p_{2}}\otimes\ket{-\frac{1}{2}})
\rightarrow \frac{1}{\sqrt{2}}(\ket{\Lambda p_{1}}-\ket{\Lambda
p_{2}})\otimes\ket{\frac{1}{2}}, \ee which is a separable state.\\

\section{Conclusions}
In this paper, we have considered spin-momentum correlation of
massive single spin-$\frac{1}{2}$ particle quantum states which furnish an
irreducible representation of the Poincare group. Instead of the superposition of all momenta we have considered only  two momenta $p_1$ and $p_2$ eigen states. We have shown that the spin-momentum
correlation of relativistic  single spin-$\frac{1}{2}$ particle
mixed state( when the momentum is perpendicular to the boost
direction) is dependent on the angle between spin and momentum and when they are parallel the measure of entanglement    decreases with increasing of velocity of the observer.
We have also shown that the Lorentz transformations can be
realized as quantum control gates and they become like the CNOT
gate in the limit where $\beta\rightarrow 1$.

 \vspace{1cm}\setcounter{section}{0}
\setcounter{equation}{0}
\renewcommand{\theequation}{A-\roman{equation}}
{\Large APPENDIX A}\\

{\bf Wigner representation for spin-$\frac{1}{2}$}

In Ref. \cite{weinberg}, is shown that effect of an arbitrary Lorentz
transformation $\Lambda$ unitarily implemented as $U(\Lambda)$ on
single-particle states is \be
\label{wig1}U(\Lambda)(\ket{p}\otimes\ket{\sigma})=\sqrt{\frac{(\Lambda
p)^0}{p^0}}\sum_{\sigma^\prime} D_{\sigma^\prime
\sigma}(W(\Lambda,p))(\ket{\Lambda p}\otimes\ket{\sigma^\prime})
,\ee where \be \label{wig11}W(\Lambda,p)=L^{-1}(\Lambda p)\Lambda
L(p),\ee\\ is the Wigner rotation \cite{Wigner}. We will consider two
reference frames in this work: one is the rest frame S and the other
is the moving frame $S^\prime$ in which a particle whose
four-momentum \emph{p} in S is seen as boosted with the velocity
$\vec{v}$. By setting the boost and particle moving directions in
the rest frame to be $\hat{v}$ with $\hat{e}$ as the normal vector
in the boost direction  and $\hat{p}$, respectively, and
$\hat{n}=\hat{e}\times \hat{p}$, the Wigner representation for
spin-$\frac{1}{2}$ particle is found as \cite{Ahn},
\be
\label{wig2}D^{\frac{1}{2}}(W(\Lambda,p)=\cos{\frac{\Omega_{\vec{p}}}{2}}+i\sin{\frac{\Omega_{\vec{p}}}{2}}(\vec{\sigma}.\hat{n}),\ee\\
where \be
\label{wig3}\cos{\frac{\Omega_{\vec{p}}}{2}}=\frac{\cosh{\frac{\alpha}{2}}\cosh{\frac{\delta}{2}}+\sinh{\frac{\alpha}{2}}\sinh{\frac{\delta}{2}}(\hat{e}.\hat{p})}{\sqrt{[\frac{1}{2}+\frac{1}{2}\cosh{\alpha}\cosh{\delta}+\frac{1}{2}\sinh{\alpha}\sinh{\delta}(\hat{e}.\hat{p})]}},\ee\\
\be
\label{wig4}\sin{\frac{\Omega_{\vec{p}}}{2}}\hat{n}=\frac{\sinh{\frac{\alpha}{2}}\sinh{\frac{\delta}{2}}(\hat{e}\times\hat{p})}{\sqrt{[\frac{1}{2}+\frac{1}{2}\cosh{\alpha}\cosh{\delta}+\frac{1}{2}\sinh{\alpha}\sinh{\delta}(\hat{e}.\hat{p})]}},\ee\\
 and
$$\cosh{\alpha}=\gamma=\frac{1}{\sqrt{1-\beta^2}} ,\cosh{\delta}=\frac{\emph{E}}{m} ,\beta=\frac{v}{c}
.$$\\

{\Large APPENDIX B}\\

{\bf Entanglement of formation}

Let $\ket{\psi}=\sum_{i,j=1}^{N} a_{ij} e_i \otimes e_j ,\ \ a_{ij}\in C$ be an two-particle pure states
with normalization $\sum_{i,j=1}^{N}|a_{ij}|^2=1$. For this
pure state the entanglement of formation \emph{E} is defined as the
entropy of either of the two sub-Hilbert space, i. e.
\be\label{e1}\emph{E}(\ket{\psi})=-Tr(\rho_{1}
\log_{2}\rho_{1})=-Tr(\rho_{2} \log_{2}\rho_{2}).\ee where
$\rho_1$(respectively, $\rho_2$) is the partial  trace of
$\ket{\psi}\bra{\psi}$ over the first (respectively, second)
Hilbert space. A given density matrix
$\rho$ on $\mathcal{H}^{d}\otimes \mathcal{H}^{d}$ has pure-state
decompositions of $\ket{\psi_{i}}$ of the form (\ref{ro1}) with
probabilities $P_{i}$,  The entanglement of formation for the mixed
state $\rho$ is defined as the average entanglement of the pure
states of the decomposition,
 minimized over all possible decompositions of $\rho$, i. e. \be
\label{e3}\emph{E}(\rho)=min \sum_{i} P_{i}
 \emph{E}(\ket{\psi_i}).\ee  In the case of n=2, (\ref{e1}) can be
written as \be\label{con5}\emph{E}(\ket{\psi})|_{n=2}
=\emph{H}(\frac{1+\sqrt{1-C^2}}{2}) ,\ee  where $\emph{H}(x)=-x
\log_{2}{x}-(1-x)\log_{2}{(1-x)}$ is binary entropy and C is called
concurrence. Thus calculation of (\ref{e3}) can be  reduced to
calculate the corresponding minimum of $$C(\rho)=min
\Sigma_{b=1}^{k}p_{b}C(\ket{\psi_{b}}).$$\par  Wootters in
\cite{Wootters2} has shown that for a 2-qubit system entanglement of
formation of a mixed state $\rho$ can be defined as
\be\label{ef1}\emph{E}(\rho) =\emph{H}(\frac{1+\sqrt{1-C^2}}{2})
,\ee by
\be\label{con3}C(\rho)=max(0,\lambda_1-\lambda_2-\lambda_3-\lambda_4),\ee
where the $\lambda_i$ are the non-negative eigenvalues, in
decreasing order, of the Hermitian matrix $$R\equiv
\sqrt{\sqrt{\rho}\tilde{\rho}\sqrt{\rho}},$$  and
$$\tilde{\rho}=(\sigma_y\otimes \sigma_y)\rho^{\ast}(\sigma_y\otimes
\sigma_y),$$ where $\rho^{\ast}$is the complex conjugate of $\rho$
when it is expressed in a fixed basis such as
${\ket{\uparrow},\ket{\downarrow}}$, and $\sigma_{y}$ is
$\begin{array}{cc}
  \left(\begin{array}{cc}
  0 & -i \\
  i & 0 \\
\end{array}
\right)\\
\end{array}
$ on the same bases.\par In order to obtain the concurrence of
\emph{BD} states (\ref{ro1}) we follow the method presented by
Wootters in \cite{Wootters2}. We define subnormalized orthogonal
eigenvectors $\ket{v_i}$ as
$$\ket{\emph{v}_i}=\sqrt{\emph{P}_i}\ket{\psi_i} , \langle{\emph{v}_i}\ket{\emph{v}_i}=\emph{P}_i \delta_{ij} ,$$
and define $\ket{\emph{x}_i}$ as $\ket{\emph{x}_i}=\Sigma_{j=1}^{4}
U_{ij}^{\ast} \ket{\emph{v}_i}$ for $i=1,2,3,4 $ such that
$$\bra{x_i}{\tilde{x_j}}\rangle=(U\tau U^{T})_{ij}=\lambda_{i}\delta_{ij} ,$$
$$\ket{\tilde{x_j}}=\sigma_y \otimes \sigma_y \ket{x^\ast_j}$$
where $\tau_{ij}=\bra{\emph{v}_{i}}{\tilde{\emph{v}_{j}}}\rangle$is
a symmetric but not necessarily Hermitian matrix. In construction of
$\ket{x_i}$ we have  considered  the fact that for any symmetric
matrix $\tau$ one can always find a unitary matrix U in such a way
that $\lambda_i$ are real and non-negative, that is, they are the
square roots of eigenvalues of $\tau \tau^{\ast}$ which are the same
as the eigenvalues of R. Moreover one can always find U such that
$\lambda_1$ being the largest one. After some calculations we get
the following values for $\lambda_i$ ,
$$\lambda_1=P_1 ,\lambda_2=P_2 ,\lambda_3=P_3 ,\lambda_4=P_4  ,$$
and concurrence can be evaluated as
\be\label{con2}C(\rho)=(P_1-P_2-P_2-P_4).\ee

\end{document}